\documentclass[aps,prl,showpacs,superscriptaddress,numerical,twocolumn,amsmath,amssymb,floatfix,reprint]
{revtex4-1}

\usepackage{graphicx}
\usepackage{dcolumn}
\usepackage{bm}
\usepackage[
	pdffitwindow=true,
	colorlinks=true,
	frenchlinks=false,
        linkcolor=blue,
	anchorcolor=blue,
        citecolor=blue,
        filecolor=blue,
        urlcolor=blue,
        bookmarks=true,
        bookmarksopen=true,
	bookmarksnumbered=true, 
        bookmarksopenlevel=1,
        plainpages=false,
	pdfpagelayout=OnePage,
        pdfpagelabels=true,
	breaklinks
]{hyperref}

\usepackage[per-mode=symbol,separate-uncertainty]{siunitx}
\usepackage[caption=false]{subfig}
\usepackage{braket}
\renewcommand\bra[1]{{\langle{#1}|}} 
\makeatletter
\renewcommand\ket[1]{%
  \@ifnextchar\bra{\k@t{#1}\!}{\k@t{#1}}%
}
\newcommand\k@t[1]{{|{#1}\rangle}}
\makeatother

\graphicspath{{Figures/}}

\begin{document}

\preprint{APS/123-QED}

\title{Qubit Measurement by Multichannel Driving}

\author{Joni Ikonen}
\affiliation{QCD Labs, QTF Centre of Excellence, Department of Applied Physics, Aalto University, P.O.~Box 13500, FI-00076 Aalto, Finland}
\author{Jan Goetz}
\affiliation{QCD Labs, QTF Centre of Excellence, Department of Applied Physics, Aalto University, P.O.~Box 13500, FI-00076 Aalto, Finland}
\author{Jesper Ilves}
\affiliation{QCD Labs, QTF Centre of Excellence, Department of Applied Physics, Aalto University, P.O.~Box 13500, FI-00076 Aalto, Finland}
\author{Aarne Ker\"anen}
\affiliation{QCD Labs, QTF Centre of Excellence, Department of Applied Physics, Aalto University, P.O.~Box 13500, FI-00076 Aalto, Finland}
\author{Andras M.\ Gunyho}
\affiliation{QCD Labs, QTF Centre of Excellence, Department of Applied Physics, Aalto University, P.O.~Box 13500, FI-00076 Aalto, Finland}
\author{Matti Partanen} 
\affiliation{QCD Labs, QTF Centre of Excellence, Department of Applied Physics, Aalto University, P.O.~Box 13500, FI-00076 Aalto, Finland}
\author{Kuan Y.\ Tan} 
\affiliation{QCD Labs, QTF Centre of Excellence, Department of Applied Physics, Aalto University, P.O.~Box 13500, FI-00076 Aalto, Finland}
\author{Dibyendu Hazra}
\affiliation{QCD Labs, QTF Centre of Excellence, Department of Applied Physics, Aalto University, P.O.~Box 13500, FI-00076 Aalto, Finland}
\author{Leif Gr\"onberg} 
\affiliation{VTT Technical Research Centre of Finland, QTF Center of Excellence, P.O.~Box 1000, FI-02044 VTT, Finland}
\author{Visa Vesterinen}
\affiliation{QCD Labs, QTF Centre of Excellence, Department of Applied Physics, Aalto University, P.O.~Box 13500, FI-00076 Aalto, Finland}
\affiliation{VTT Technical Research Centre of Finland, QTF Center of Excellence, P.O.~Box 1000, FI-02044 VTT, Finland}
\author{Slawomir Simbierowicz}
\affiliation{VTT Technical Research Centre of Finland, QTF Center of Excellence, P.O.~Box 1000, FI-02044 VTT, Finland}
\author{Juha Hassel}
\affiliation{VTT Technical Research Centre of Finland, QTF Center of Excellence, P.O.~Box 1000, FI-02044 VTT, Finland}
\author{Mikko M\"ott\"onen}
\affiliation{QCD Labs, QTF Centre of Excellence, Department of Applied Physics, Aalto University, P.O.~Box 13500, FI-00076 Aalto, Finland}

\date{\today}

\begin{abstract} 
We theoretically propose and experimentally implement a method of measuring a qubit by driving it close to the frequency of a dispersively coupled bosonic mode. 
The separation of the bosonic states corresponding to different qubit states begins essentially immediately at maximum rate, leading to a speedup in the measurement protocol. 
Also the bosonic mode can be simultaneously driven to optimize measurement speed and fidelity.
We experimentally test this measurement protocol using a superconducting qubit coupled to a resonator mode. For a certain measurement time, we observe that the conventional dispersive readout yields close to \SI{100}{\percent} higher average measurement error than our protocol. 
Finally, we use an additional resonator drive to leave the resonator state to vacuum if the qubit is in the ground state during the measurement protocol. 
This suggests that the proposed measurement technique may become useful in unconditionally resetting the resonator to a vacuum state after the measurement pulse.
\end{abstract}

\maketitle

Since the birth of quantum mechanics, quantum measurements and the related collapse of the wavefunction has puzzled scientists~\cite{Bassi_2013,Fuwa_2015}, culminating in various interpretations of quantum mechanics such as that of many worlds~\cite{DeWitt_2015}. With the recent rise of quantum technology~\cite{OBrien_2009,Ladd_2010,Kurizki_2015}, the quantum measurement has become one of the most important assets for practical applications. For example, measurements of single qubits are the key in reading out the results of quantum computations~\cite{Riste_2015,Kelly_2015,Hacohen_2016,Reagor_2018} and parity measurements in multi-qubit systems are frequently required in quantum error correction codes such as the surface and color codes~\cite{Fowler_2009,Fowler_2011,Riste_2013,Barends_2014,Nigg_2014}. Furthermore, single-qubit measurement and feedback can be used to reset qubits~\cite{Geerlings_2013a,Bultink_2016,Magnard_2018} or even solely provide the non-linearity needed to implement multi-qubit gates~\cite{Knill_2001,Nielsen_2004,Kok_2007}.

One of the most widespread ways to measure qubits is to couple them to one or several bosonic modes, such as those of the electromagnetic field, and to measure their effect on the radiation~\cite{Gambetta_2007}. This method is currently used, for example, in quantum processors based on superconducting circuits~\cite{Riste_2012,Jeffrey_2014,Goetz_2016a,Walter_2017,Weber_2017,Rosenblum_2018}, quantum dots~\cite{Colless_2013,Scarlino_2017,Mi_2017}, and trapped ions~\cite{Blatt_2012}. Especially with the rise of circuit quantum electrodynamics~\cite{Blais_2004,Wallraff_2004}, this measurement technique has become available to many different hybrid systems such as mechanical oscillators~\cite{Bochmann_2013,Pirkkalainen_2013} and magnons~\cite{Tabuchi_2015}.

Theoretically, the interacting system of a qubit and a bosonic mode is surprisingly well described by the Jaynes--Cummings model~\cite{Jaynes_1963,Shore_1993}. If the qubit frequency is far detuned from the mode frequency, i.e., we operate in the dispersive regime, the interaction term renders the mode frequency to depend on the qubit state. Consequently, a straightforward way to implement a non-demolition measurement on the qubit state is to drive the mode at a certain frequency close to the resonance and measure the phase shift of the output field with respect to the driving field. This kind of dispersive measurement has been extremely successful, for example, in superconducting qubits~\cite{Wallraff_2005} with increasing accuracy and speed~\cite{Mallet_2009,Jerger_2012,Jeffrey_2014,Heinsoo_2018} currently culminating in \SI{99.2}{\percent} fidelity in \SI{88}{\nano\second}~\cite{Walter_2017}. 

In the dispersive measurement, one of the key issues has been the ability to quickly populate the bosonic mode in the beginning of the measurement protocol~\cite{Jeffrey_2014} without surpassing the critical photon number, and to quickly evacuate the excitations from the mode after the measurement~\cite{Bultink_2016,McClure_2016}. These requirements point to the need for a fast, low-quality readout mode. However, this poses a trade-off on the qubit lifetime, which to some extent, can be answered using Purcell filters~\cite{Jeffrey_2014,Bronn_2015,Walter_2017} with the cost of added circuit complexity. A simple and fast high-fidelity measurement scheme is of great interest not only to the field of superconducting qubits, but also to other quantum technology platforms utilizing bosonic modes as the measurement tool.

Inspired by our recent work~\cite{Ikonen2} on quickly stabilizing resonator states by a qubit drive, we propose in this letter a qubit measurement protocol which is based on driving the qubit close to the frequency of the bosonic mode through a non-resonant channel. Owing to the dispersive coupling, the initial vacuum state of the resonator begins to rotate selectively on the qubit state about a point controlled by the strength and phase of the qubit drive. Importantly, this rotation begins immediately after the drive pulse arrives at the qubit with no bandwidth limitation imposed by the resonator. We demonstrate this non-demolition readout scheme in planar superconducting qubits and observe that it leads to a significant speedup. Furthermore, we discuss how this method can be used to unconditionally reset the resonator state into vacuum after the readout without the need for feedback control. We experimentally demonstrate a related effect where the resonator is left in the vacuum state provided that the qubit was in the ground state.

\begin{figure}
  \centering
  \includegraphics{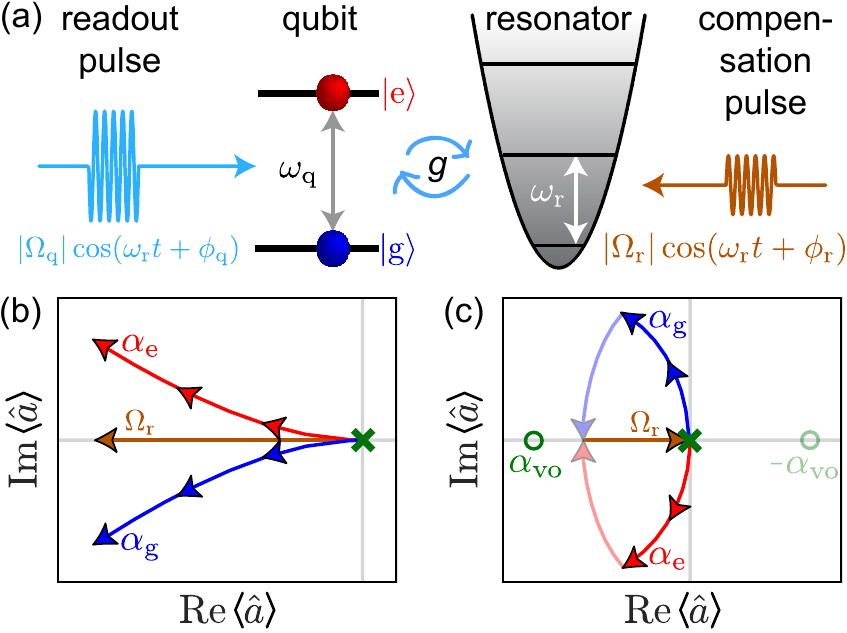} 
  \caption{\label{fig:1} (a) Schematic presentation of the readout scheme where the qubit is driven (blue color) at the frequency of a dispersively coupled bosonic mode. A compensation tone (brown color) on the resonator may be used to optimize the result. We consider the case where the detuning $\Delta\,{=}\,\omega_\textrm{q}\,{-}\,\omega_\textrm{r}$ is much greater than the qubit--mode coupling strength $g$. (b) Evolution of the mean of the resonator state in phase space for the conventional dispersive readout starting from vacuum (cross) provided that the qubit was prepared in $|\textrm{g}\rangle$ (blue) or $|\textrm{e}\rangle$ (red). (c) As (b) but the readout pulse is applied directly to the qubit. Thus, the resonator states start to rotate about a new virtual origin $\alpha_{\textrm{vo}}$ (circle) leading to a faster separation. After measurement, we may reverse the sign of the virtual origin and wait for the resonator sates corresponding to different qubit sates to fully overlap (faint colors). A subsequent shift (brown color) finalizes an unconditional reset of the resonator.}
\end{figure}   

Let us theoretically study a qubit coupled to a single bosonic mode such as that of an electromagnetic resonator, as shown in Fig.~\ref{fig:1}(a). Instead of using the conventional readout by populating the resonator with a coherent pulse~\cite{Wallraff_2005,Mallet_2009}, we measure the qubit state by driving the qubit at the resonator frequency $\omega_{\textrm{r}}$. 
In addition, we can apply a compensation pulse to the resonator to eliminate cross-coupling effects with the qubit or otherwise control the resonator state. The qubit and the resonator couple to their respective driving fields with different strengths, which together with the pulse envelopes constitute the effective Rabi angular frequencies $\Omega_{\textrm{q}}$ and $\Omega_{\textrm{r}}$, respectively. 
The qubit may be first excited from the ground state $\ket{\textrm{g}}$ to the excited state $\ket{\textrm{e}}$ by a separate drive tone at the transition angular frequency $\omega_{\textrm{q}}\,{=}\,\omega_{\textrm{r}}\,{+}\,\Delta$, where $\Delta$ is the detuning.

In the frame rotating at $\omega_{\textrm{r}}$ with respect to the bare qubit and resonator Hamiltonians, $\hbar\omega_{\textrm{q}}\hat{\sigma}_{+}\hat{\sigma}_{-}$ and $\hbar\omega_{\textrm{r}}\hat{a}^{\dagger}\hat{a}$, respectively, the system is described by the Jaynes--Cummings Hamiltonian 
\begin{equation}
\hat{H}/\hbar=\Delta\hat{\sigma}_{+}\hat{\sigma}_{-}+(g\hat{\sigma}_{+}\hat{a}+\Omega_{\textrm{q}}\hat{\sigma}_{+}+i\Omega_{\textrm{r}}\hat{a}^{\dagger}+\textrm{H.c.}),\label{eq:Hamilton1}
\end{equation}
where $g$ denotes the qubit--resonator coupling strength, $\hat{a}^{\dagger}$ and $\hat{\sigma}_{+}\equiv\ket{\textrm{e}}\bra{\textrm{g}}$ are the creation operators of the resonator mode and of the qubit, respectively.
Above, we have introduced the rotating-wave approximation justified by $g\,{\ll}\,\omega_{\textrm{q}},\omega_{\textrm{r}}$.

To demonstrate the benefit of driving the qubit at the frequency of the resonator, we employ the standard dispersive approximation~\cite{Boissonneault_2009} in the regime $g\,{\ll}\,\left|\Delta\right|$. This yields, up to constant energy terms, the Hamiltonian~\cite{supp}
\begin{eqnarray}
\hat{H}^{\prime\prime}/\hbar & \approx & \left(\Delta+\chi\right)\hat{\sigma}_{+}\hat{\sigma}_{-}+\left[\left(\Omega_{\text{q}}+i\Omega_{\mathrm{r}}\frac{\chi}{g}\right)\hat{\sigma}_{+}+\text{H.c.}\right]\nonumber \\
 &  & -\chi\hat{\sigma}_{z}\hat{a}^{\dagger}\hat{a}+\left[\left(i\Omega_{\text{r}}-\Omega_{\text{q}}\frac{\chi}{g}\hat{\sigma}_{\text{z}}\right)\hat{a}^{\dagger}+\text{H.c.}\right], \label{eq:hamilton2}
\end{eqnarray}
where $\chi\,{=}\,g^{2}/\Delta$ is the dispersive shift for a two-level system and $\hat{\sigma}_\text{z}\,{=}\,|\text{g}\rangle\langle\text{g}|\,{-}\,|\text{e}\rangle\langle\text{e}|$. The term proportional to $\hat{a}^{\dagger}$ is a generator of a displacement operator that depends on the state of the qubit. 
Thus, driving the qubit effectively realizes longitudinal coupling \cite{Didier2015, Billangeon2015} for the duration of the readout, implying that the rate of state separation is not limited by the rate at which the resonator is populated.

In our work, the resonator is accurately described by a coherent state $\ket{\alpha}$ such that $\hat{a}\ket{\alpha}\,{=}\,\alpha\ket{\alpha},\;\alpha\,{\in}\,\mathbb{C}$. The drive amplitude $\Omega_{\text{q}}$ may be turned on very fast, causing the amplitudes $\alpha_{\text{g/e}}$ corresponding to the eigenstates of the qubit, $\ket{\textrm{g}}$ and $\ket{\textrm{e}}$, to separate in the complex plane at least with the initial speed $2\,\Omega_{\textrm{q}}\chi/g$. 
This minimum speed is achieved with $\Omega_{\text{r}}\,{=}\,0$ for an initial vacuum state in the resonator. 

As the resonator becomes populated, the trajectories begin to curve due to the dispersive term $-\chi\hat{\sigma}_{z}\hat{a}^{\dagger}\hat{a}$ in Eq.~\eqref{eq:hamilton2} and, in fact, to rotate about the point $\alpha_{\text{vo}}\,{\equiv}\,{-}\Omega_{\text{q}}/g$.
This behavior is intuitively understood in a frame displaced by $\alpha_{\text{vo}}$.
Introducing a shifted annihilation operator $\hat{b}\,{=}\,\hat{a}\,{-}\,\alpha_{\text{vo}}$, the last line of Eq.~\eqref{eq:hamilton2} yields 
\begin{eqnarray}
\hat{H}_{\text{r}}^{\prime\prime}/\hbar & \approx & -\chi\hat{\sigma}_{z}\hat{b}^{\dagger}\hat{b}+\left(i\Omega_{\text{r}}\hat{b}^{\dagger}+\text{H.c.}\right). \label{eq:hamilton3}
\end{eqnarray}
The first term in Eq.~\eqref{eq:hamilton3} corresponds to a rotation of the amplitude $\alpha$ in the complex plane about the virtual origin $\alpha_{\textrm{vo}}$ with an angular frequency $\chi$ in
a direction determined by the qubit state. 
Thus driving the qubit at the resonator frequency $\omega_{\textrm{r}}$ effectively shifts
the origin of the resonator phase space to a point $\alpha_{\textrm{vo}}$ in the rotating frame. 

The term $\left(\Omega_{\text{q}}\,{+}\,i\Omega_{\mathrm{r}}\chi/g\right)\hat{\sigma}_{+}$
in Eq.~\eqref{eq:hamilton2} shows that the drives slightly tilt the qubit Hamiltonian. The tilt of the quantization axis determines the speed at which the drives can be turned on while maintaining adiabaticity, the lowest-order condition being approximately $\dot{\Omega}_{\textrm{q}}\,{\ll}\,\Delta^{2}/\sqrt{2}$.
Since $\Omega_{\textrm{q}}\,{\ll}\,\Delta$, the rise time of the qubit drive pulse can be negligibly short compared with the relevant dynamics of the resonator states. Thus the qubit-state-dependent separation dynamics of the resonator state starts to take place essentially instantly in this readout protocol.

In contrast to the multichannel readout visualized in Fig.~\ref{fig:1}(c), the usual dispersive readout relies solely on the term $-\chi\hat{\sigma}_{z}\hat{a}^{\dagger}\hat{a}$, which implies that one needs to use the resonator drive to populate the resonator for the state separation to take place, see Fig.~\ref{fig:1}(b). 
The characteristic time scale for the population dynamics $1/\kappa$ is determined by the internal and external damping rates of the resonator $\kappa_{\textrm{i}}$ and $\kappa_{\textrm{x}}$, respectively, as $\kappa=\kappa_{\textrm{i}}+\kappa_{\textrm{x}}$. 

In addition to the potentially faster readout, our scheme offers more control over the evolution of the states than the usual dispersive readout. For example, we may continuously drive the resonator such that either $\alpha_{\textrm{e}}$ or $\alpha_{\textrm{g}}$ end in any desired position at the end of the readout.
For example, choosing $i\Omega_{\mathrm{r}}\,{=}\,\Omega_{\text{q}}\chi/g$ in Eq.~\eqref{eq:hamilton2} causes $\alpha_{\textrm{g}}$ to remain in vacuum while $\alpha_{\textrm{e}}$ is displaced. Interestingly, we may also reset the resonator to the vacuum state unconditionally
on the qubit state and without feedback control. 
As illustrated in Fig.~\ref{fig:1}(c), one may shift the phase of $\alpha_{\textrm{vo}}$ by $\pi$ after the actual measurement pulse and wait for both of the amplitudes $\alpha_{\textrm{e}}$ and $\alpha_{\textrm{g}}$ to rotate on top of each other. Subsequently, both distributions may be shifted to the vacuum state using a single pulse on the resonator.

Note that due to the finite resonator bandwidth, the resonator will slowly saturate towards a steady state. 
We obtain the steady states by solving the standard Lindblad master equation $\dot{\hat{\rho}}\,{=}\,{-}i[\hat{H},\hat{\rho}]/\hbar\,{+}\,\kappa\mathcal{L}[\hat{a}]\hat{\rho}/2$, where $\mathcal{L}[\hat{a}]$ is the Lindblad superoperator and $\hat{\rho}$ is the density operator of the qubit--resonator system. Forcing the states to remain coherent, the steady states $|\alpha^\text{s}_\textrm{g/e}\rangle$ are given by
$\alpha^\text{s}_\textrm{g/e}\,{=}\,(i\Omega_\textrm{r}\,{\mp}\,\Omega_\textrm{q}\chi/g)/(i\kappa/2\,{\pm}\,\chi)$.  Above, we have restricted our theory to the case of a two-level system. However, the scheme also works in the case of many non-equidistant levels~\cite{supp} such as those of a superconducting transmon qubit~\cite{Koch_2007} studied below. Here, the driving frequency needs to be slightly offset from that of the resonator and an additional resonator drive is needed to obtain essentially Eq.~\eqref{eq:hamilton2} for the transmon. Note that qubit non-linearity is pivotal to obtain a non-vanishing dispersive shift $\chi$.

\begin{figure}
  \centering
  \includegraphics{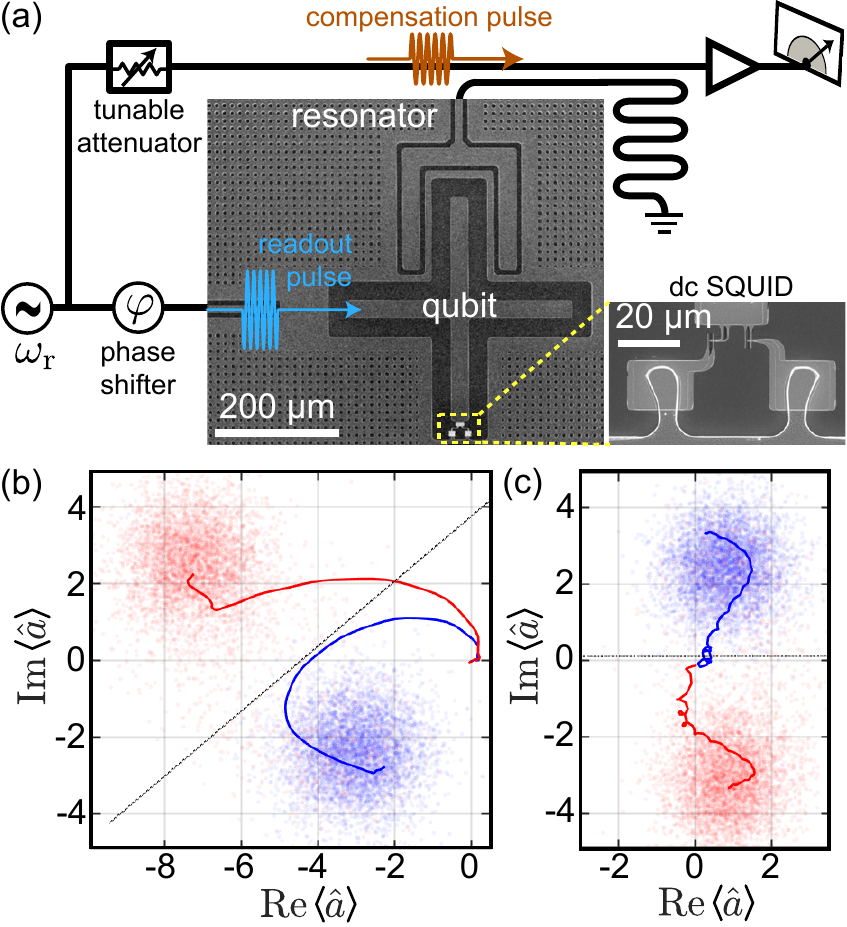} 
  \caption{\label{fig:2} (a) Simplified measurement setup and scanning electron micrographs of the experimental sample realizing the theoretical scheme. We employ an Xmon qubit~\cite{Barends_2013}, the frequency of which may be tuned by applying an external magnetic flux to the accompanying dc~SQUID. The qubit is capacitively coupled to a voltage drive line and to a co-planar waveguide resonator which is, in turn, coupled to a transmission line. After amplification, we measure the two quadratures $\textrm{Re}\,\hat{a}$ and $\textrm{Im}\,\hat{a}$ of the resonator field. (b) Evolution of the amplitudes of the coherent states corresponding to the ground (blue line) and excited (red line) states of the qubit during a 420-ns conventional dispersive readout.
  Corresponding results of single-shot measurements are shown by dots (see text for details). The dotted line indicates the threshold for assigning the measurement outcome.
   (c) As (b), but the drive tone is applied to both the qubit and the resonator with an optimized relative phase. The single-shot measurement fidelities are \SI{96.4}{\percent} and  \SI{96.6}{\percent} for (b) and (c), respectively. }
\end{figure}

To implement our theoretical scheme we have fabricated~\cite{supp} a superconducting Xmon qubit~\cite{Barends_2013} shown in Fig.~\ref{fig:2}(a). It is coupled with strength $g\,{=}\,2\pi{\times}\SI{130}{\mega\hertz}$ to a coplanar waveguide resonator of frequency $\omega_{\textrm{r}}/2\pi\,{=}\,\SI{6.02}{\giga\hertz}$. The resonator has internal and external loss rates $\kappa_{\textrm{i}}\,{=}\,2\pi{\times}\SI{0.5}{\mega\hertz}$ and $\kappa_{\textrm{x}}\,{=}\,2\pi{\times}\SI{1.5}{\mega\hertz}$, respectively. 
We tune the qubit to the point of optimal phase coherence~\cite{supp}, $\omega_{\textrm{q}}/2\pi\,{=}\,\SI{7.86}{\giga\hertz}$, where it is characterized by the energy relaxation time $T_{1}\,{=}\,\SI{3.0}{\micro\second}$. This leads to a dispersive shift $\chi\,{=}\,-2\pi{\times}\SI{1.6}{\mega\hertz}$. We mount the sample to the base temperature stage, $T\,{=}\,\SI{20}{\milli\kelvin}$, of a dilution refrigerator and extract the effective qubit temperature $T_{\textrm{eff}}\,{=}\,\SI{73}{\milli\kelvin}$ from histograms of single-shot measurements~\cite{supp}. For this purpose, we use a traveling-wave parametric amplifier~\cite{Macklin_2015} and a heterodyne detection setup to measure the two quadratures 
$\textrm{Re}\,\hat{a}$ and $\textrm{Im}\,\hat{a}$ 
of the resonator field.  

Figures~\ref{fig:2}(b) and~\ref{fig:2}(c) present the experimentally measured temporal trajectories of ensemble-averaged expectation values $\alpha(t)\,{=}\,\langle \hat{a} (t)\rangle$ for the conventional readout and our method, respectively. The trajectories show qualitative agreement with our theory: In the conventional readout, the states move in the general direction of the drive and separate as the distance to the origin increases. In our scheme, the states move to opposite directions owing to precession about the virtual origin lying on the negative real axis. The dominating differences between Figs.\ref{fig:1}(b,c) and~\ref{fig:2}(b,c) can be explained by the higher levels of the transmon~\cite{supp}.

To characterize the performance of our method, we implement single-shot measurements $S$, of the observable $ \hat{S}\,{=}\,\int_{0}^{\tau} \left[W_{\textrm{re}}(t)\textrm{Re}\,\hat{a}(t)\,{+}\,iW_{\textrm{im}}(t)\textrm{Im}\,\hat{a}(t)\right] \textrm{d}t$ by temporal integration of the readout signal. Here, the normalized weight functions are determined from the previously measured trajectories as $W_{\textrm{re}}(t)\,{\propto}\,\left|\textrm{Re}[\alpha_\textrm{e}(t)\,{-}\,\alpha_\textrm{g}(t)]\right|$ and $W_{\textrm{im}}(t)\,{\propto}\,\left|\textrm{Im}[\alpha_\textrm{e}(t)\,{-}\,\alpha_\textrm{g}(t)]\right|$. Thus the most weight is given to the signal when the state separation is known to be the largest.  We also determine reference points $\alpha^\textrm{ref}_j$ by averaging shots conditioned on the qubit being in state $j\,{\in}\,{\textrm\{g,e\}}$. For a single measurement shot $S$, we infer that the qubit was in state $\ket{\textrm{g}}$ if $|S\,{-}\,\alpha^\textrm{ref}_\textrm{g}|{\,<\,}|S\,{-}\,\alpha^\textrm{ref}_\textrm{e}|$, see Figs.~\ref{fig:2}(b) and~\ref{fig:2}(c).

The error probability of assigning an incorrect label for the intended qubit state is calculated as $\epsilon_\textrm{total}\,{=}\,[p(\textrm{e}\,|\,\textrm{g})\,{+}\,p(\textrm{g}\,|\,\textrm{e})]/2$, where $p({j\,|\,k})$ is the sampled probability to assign the label $j$ to a state supposedly prepared in $\ket{k}$.
To extract the error due to readout, we independently measure the state preparation errors caused by faulty gate operations, spontaneous decay, and thermal excitation. We estimate that these sources account for  $\epsilon_{\textrm{prep}}\,{=}\,\SI{2.6}{\percent}$ of the total error, mainly limited by $T_{1}$ decay of our sample (see~\cite{supp} for details).

\begin{figure}
  \centering
  \includegraphics{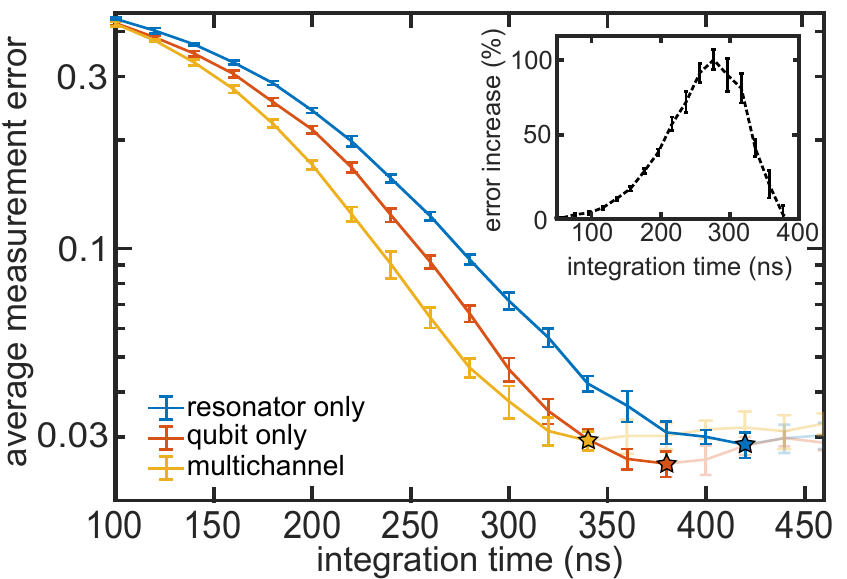} 
  \caption{\label{fig:3} Average measurement error $\epsilon_\textrm{total}\,{-}\,\epsilon_\textrm{prep}$ as a function of integration time for the conventional readout (blue markers), qubit driving (red markers), and multichannel driving (yellow markers). Each data point shows the average and the standard deviation of 10 measurement runs consisting of $10^4$ single-shot measurements. The stars indicate the time for which the lowest error is obtained for each method. The inset shows the relative increase in the measurement error when the readout method is changed from the multichannel scheme to the conventional readout. When we drive the resonator only, the experimental parameters are identical to those in Fig.~\ref{fig:2}(b) and with multichannel driving to those in Fig.~\ref{fig:2}(c). For the multichannel readout, the drive powers to the resonator and to the qubit are decreased by \SI{2}{\decibel} and \SI{1}{\decibel} compared with single-channel driving, respectively. }
\end{figure}

We benchmark the speed and fidelity of our readout scheme against the conventional method in Fig.~\ref{fig:3}, which demonstrates that driving the qubit directly, with or without the compensation tone on the resonator, yields considerably lower errors for integration times $\tau\,{\leq}\,\SI{350}{\nano\second}$. Thus, measuring the qubit state by direct or multichannel driving results in a noticeable speedup over driving only the resonator. For each readout scheme, the drive power is independently maximized with the condition that the third level of the transmon is negligibly excited during readout, to ensure that the readout realizes a non-demolition measurement. For the multichannel readout, the relative phase between the resonator and qubit drives $\phi_{\textrm{r}}\,{-}\,\phi_{\textrm{q}}$ is also optimized to achieve the fastest decrease in error. We observe that for a given integration time, the conventional readout bears close to \SI{100}{\percent} larger measurement error than the multichannel driving scheme.

\begin{figure}
  \centering
  \includegraphics{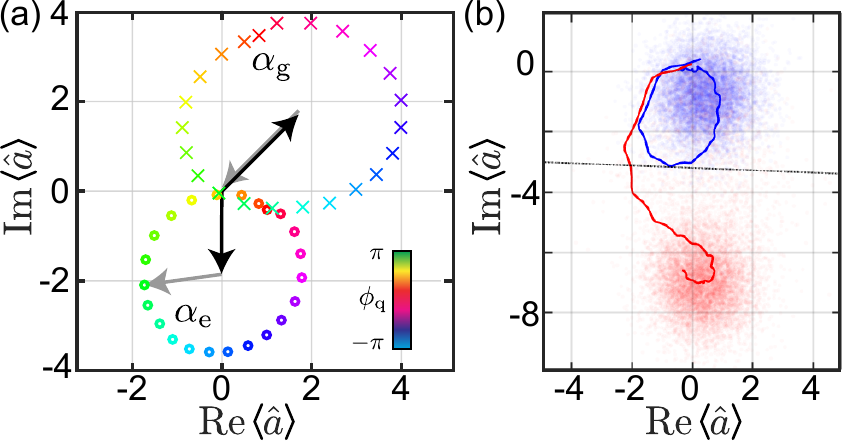} 
  \caption{\label{fig:4} (a) Measured means of the steady states corresponding to $|\textrm{g}\rangle$ ($\alpha_\textrm{g}$) and $|\textrm{e}\rangle$ ($\alpha_\textrm{e}$) in the multichannel readout as functions of the phase $\phi_{\textrm{q}}$ of the qubit drive pulse, as indicated by the different colors of the markers. The black arrows denote phase-independent contributions to the steady state owing to the resonator drive. With a particular choice of $\phi_{\textrm{q}}$, indicated by the gray arrows, one of the resonator states returns to vacuum during the measurement. (b) Evolution of the amplitude of the coherent state (solid lines) for qubit ground (blue color) and excited (red color) states in the partial reset scheme.
  The phase  $\phi_{\textrm{q}}$ is chosen such that the steady state for $|\textrm{g}\rangle$ lies at the origin. 
  }
\end{figure}  

 Combining the two drive channels provides versatile tools for controlling the state of the resonator. In Fig.~\ref{fig:4}(a), we show that as a function of the phase $\phi_{\mathrm{q}}$, the steady states draw circles in the complex plane, the centers and radii of which depend on the qubit state. This behavior is in agreement with the above result $\alpha^\text{s}_\textrm{g/e}\,{=}\,(i\Omega_\textrm{r}\,{\mp}\,\Omega_\textrm{q}\chi/g)/(i\kappa/2\,{\pm}\,\chi)$. It appears possible to choose the phase of $\Omega_\textrm{q}$ such that the resonator state corresponding to one of the qubit states remains in the vacuum state (grey arrows), a situation inaccessible by driving only the resonator. In Fig.~\ref{fig:4}(b), we show that with the multichannel method we can indeed leave $\alpha_\text{g}$ near the origin while significantly displacing $\alpha_\text{e}$. As discussed above, a related mechanism may be utilized to unconditionally reset the resonator after the readout to further reduce the duration of the overall measurement protocol.

In conclusion, we have proposed and experimentally demonstrated a readout scheme for a qubit dispersively coupled to a bosonic mode. By driving both the qubit and the mode close to the mode frequency, the readout can be turned on much faster than any other relevant time scale in the system and the resonator can be unconditionally brought back to the vacuum state without the need for feedback control. Our experiments with a superconducting qubit demonstrate  resonator control through the qubit. For a given readout time in our sample, we experimentally observe that the conventional readout may lead to more than 100\% larger error than that of the proposed scheme.   

In the future, we aim to realize the unconditional reset protocol and to optimize the sample design such that we improve on the present state-of-the-art readout~\cite{Walter_2017}. Furthermore, our proposal could be implemented in a variety of systems such as qubits coupled to nanomechanical resonators~\cite{Bochmann_2013,Pirkkalainen_2013}. We expect that in addition to qubit readout, an extension of our protocol may also be beneficial for resonator state control such as the creation of cat states~\cite{Vlastakis}.

\begin{acknowledgments}

\emph{Acknowledgments---}We acknowledge William Oliver, Greg Calusine, Kevin O'Brien, and Irfan Siddiqi for providing us with the traveling-wave parametric amplifier used in the experiments. This research was financially supported by European Research Council under Grant No.~681311 (QUESS) and Marie Sk\l{}odowska-Curie Grant No.~795159; by Academy of Finland under its Centres of Excellence Program grants Nos.~312300, 312059 and other grants Nos.~265675, 305237, 305306, 308161, 312300, 314302, 316551, and 319579; Finnish Cultural Foundation, the Jane and Aatos Erkko Foundation, Vilho, Yrj\"{o} and Kalle V\"{a}is\"{a}l\"{a} Foundation,  and the Technology Industries of Finland Centennial Foundation. We thank the provision of facilities and technical support by Aalto University at OtaNano~\---~Micronova Nanofabrication Centre.
\end{acknowledgments}

\emph{Note added: At the final stages of our work, a preprint~\cite{Touzard_2018} pursuing a similar readout scheme came to our attention. Our work is fully independent of this reference.}

\bibliographystyle{apsrev4-1}
\bibliography{Refs}

\end{document}


\preprint{AIP/123-QED}

\title{Supplemental Materials: Qubit Measurement by Multi-Channel Driving}

\author{Joni Ikonen}
\affiliation{QCD Labs, QTF Centre of Excellence, Department of Applied Physics, Aalto University, P.O.~Box 13500, FI-00076 Aalto, Finland}
\author{Jan Goetz}
\affiliation{QCD Labs, QTF Centre of Excellence, Department of Applied Physics, Aalto University, P.O.~Box 13500, FI-00076 Aalto, Finland}
\author{Jesper Ilves}
\affiliation{QCD Labs, QTF Centre of Excellence, Department of Applied Physics, Aalto University, P.O.~Box 13500, FI-00076 Aalto, Finland}
\author{Aarne Ker\"anen}
\affiliation{QCD Labs, QTF Centre of Excellence, Department of Applied Physics, Aalto University, P.O.~Box 13500, FI-00076 Aalto, Finland}
\author{Andras M.\ Gunyho}
\affiliation{QCD Labs, QTF Centre of Excellence, Department of Applied Physics, Aalto University, P.O.~Box 13500, FI-00076 Aalto, Finland}
\author{Matti Partanen} 
\affiliation{QCD Labs, QTF Centre of Excellence, Department of Applied Physics, Aalto University, P.O.~Box 13500, FI-00076 Aalto, Finland}
\author{Kuan Y.\ Tan} 
\affiliation{QCD Labs, QTF Centre of Excellence, Department of Applied Physics, Aalto University, P.O.~Box 13500, FI-00076 Aalto, Finland}
\author{Dibyendu Hazra}
\affiliation{QCD Labs, QTF Centre of Excellence, Department of Applied Physics, Aalto University, P.O.~Box 13500, FI-00076 Aalto, Finland}
\author{Leif Gr\"onberg} 
\affiliation{VTT Technical Research Centre of Finland, QTF Center of Excellence, P.O.~Box 1000, FI-02044 VTT, Finland}
\author{Visa Vesterinen}
\affiliation{QCD Labs, QTF Centre of Excellence, Department of Applied Physics, Aalto University, P.O.~Box 13500, FI-00076 Aalto, Finland}
\affiliation{VTT Technical Research Centre of Finland, QTF Center of Excellence, P.O.~Box 1000, FI-02044 VTT, Finland}
\author{Slawomir Simbierowicz}
\affiliation{VTT Technical Research Centre of Finland, QTF Center of Excellence, P.O.~Box 1000, FI-02044 VTT, Finland}
\author{Juha Hassel}
\affiliation{VTT Technical Research Centre of Finland, QTF Center of Excellence, P.O.~Box 1000, FI-02044 VTT, Finland}
\author{Mikko M\"ott\"onen}
\affiliation{QCD Labs, QTF Centre of Excellence, Department of Applied Physics, Aalto University, P.O.~Box 13500, FI-00076 Aalto, Finland}

\date{\today}
%
\maketitle

\widetext
\setcounter{equation}{0}
\setcounter{figure}{0}
\setcounter{table}{0}
\makeatletter
\renewcommand{\theequation}{S\arabic{equation}}
\renewcommand{\thefigure}{S\arabic{figure}}


\section{Sample fabrication}
\label{sec:fab}

\begin{figure}[h!]
 \centering
 \includegraphics{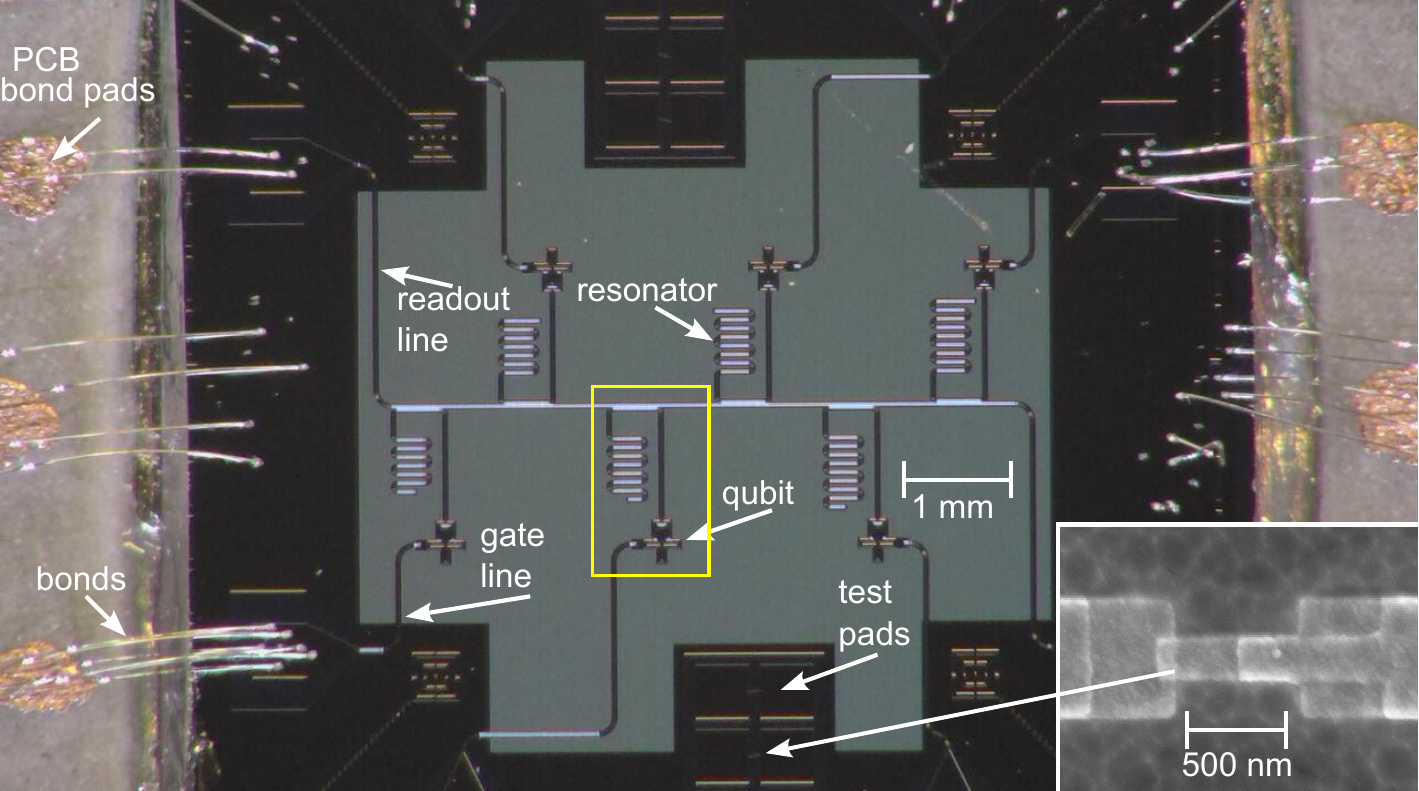}
 \caption{Photograph of the sample that we connect to a printed circuit board (PCB) with aluminum bonds. The sample contains six qubit-resonator systems and we use the highlighted qubit in our experiments. The meandering quarter-wavelength resonators are coupled to the Xmon qubits through a horseshoe-shaped capacitor. The resonators operate at different frequencies separated by \SI{200}{\mega\hertz} to enable multiplexed readout from a common transmission line. There is a voltage gate line for each qubit to control the qubit states individually. At the corner of the sample, we use test pads to characterize the resistance and size of test Josephson junctions. The grey area in the center of the sample is covered with holes in the ground plane, whereas the outer area is completely covered with niobium. The inset shows a typical Josephson junction used for the transmon qubits.}
 \label{fig:Fig_S01}
\end{figure}

We fabricate the sample shown in Fig.\,\ref{fig:Fig_S01} on a high-resistivity $(>\SI{10}{\kilo\ohm\centi\meter})$ silicon substrate. The substrate has lateral dimensions of $\SI{10}{\milli\meter}\,{\times}\,\SI{10}{\milli\meter}$ and a thickness of \SI{525}{\micro\meter}. To reduce losses due to two-level fluctuators~\cite{Sage_2011}, we remove the native oxide with argon ion milling before niobium metalization. By sputter deposition, we add \SI{200}{\nano\meter} of niobium, which we use for defining the co-planar waveguide structures. They are patterned with optical lithography and reactive ion etching. We cover all relevant sample areas with circular \SI{6}{\micro\meter} wide holes to trap flux vortices. In a next step, we define the loops for the superconducting quantum interference devices (SQUIDs) and the Josephson junctions for the transmon qubits~\cite{Koch_2007}. A double-layer Polymethyl methacrylate (PMMA) resist is patterned with electron beam lithography. The junctions have a size of $\SI{100}{\nano\meter}\,{\times}\,\SI{150}{\nano\meter}$ and are fabricated with aluminum shadow evaporation~\cite{Dolan_1977}. To reduce the loss at the aluminum--niobium interfaces~\cite{Goetz_2016}, we clean the niobium surface with argon ion milling before we start the aluminum evaporation. The junctions have a resistance of \SI{8}{\kilo\ohm} each, resulting in a Josephson energy $E_{\mathrm{J}}\,{=}\,h\,{\times}\,\SI{34}{\giga\hertz}$. The transmon qubit is made in the Xmon design~\cite{Barends_2013} and has a charging energy $E_{\mathrm{c}}\,{=}\,h\,{\times}\,\SI{264}{\mega\hertz}$. We mount the sample chip in a gold plated sample box made from copper and connect the feedlines to a printed circuit board (PCB) with aluminum wire bonds.

\begin{figure}[t]
 \centering
 \includegraphics[width=\textwidth]{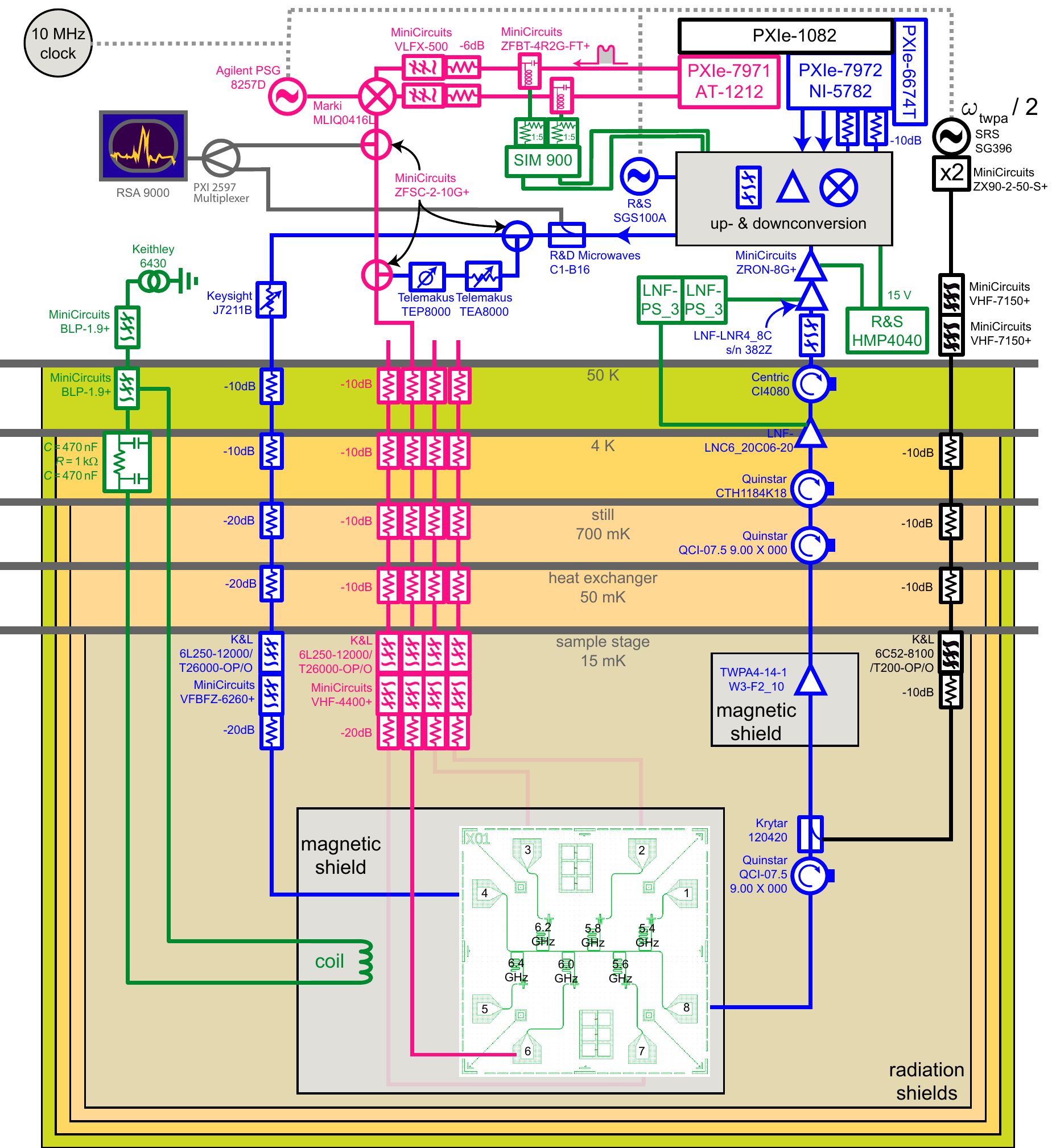}
 \caption{Measurement setup: We use a heterodyne detection scheme (upper part of the figure) to characterize the qubit-state-dependent shift of superconducting resonators mounted in a dilution refrigerator (lower part of the figure). Details are described in the text. Up and down conversion is depicted in Fig.\,\ref{fig:Fig_S02a}}
 \label{fig:Fig_S02}
\end{figure}

\section{Measurement setup}
\label{sec:setup}

\textbf{Cryogenic setup}---For our experiments, we mount the sample to the base temperature of \SI{20}{\milli\kelvin} of a dilution refrigerator (see Fig.\,\ref{fig:Fig_S02}). Our low-temperature setup employs multistage shielding against magnetic flux noise and thermal radiation containing a $\mu$-metal shield and an aluminum shield at the sample stage. In addition, we use a $\mu$-metal shield to protect a Josephson traveling-wave parametric amplifier (TWPA)~\cite{Macklin_2015} against stray magnetic fields. This amplifier is required for the single-shot measurements and has been fabricated at MIT Lincoln Laboratory. We operate the TWPA with a coherent drive that we optimize for a minimum noise temperature of the TWPA. Using an automated optimization protocol, we find a TWPA noise temperature of approximately \SI{380}{\milli\kelvin} for a gain of \SI{21.4}{\decibel} at \SI{6.2}{\giga\hertz}. To obtain the TWPA noise temperature, we have used an in-situ power calibration based on the photon-number-dependent frequency shift of the qubits. In addition to the TWPA, we use a high-electron-mobility transistor (HEMT) amplifier at the 4-K 
stage of the cryostat. The readout and the qubit gate lines are heavily attenuated and we mount microwave filters at the sample stage to reduce the influence of thermal noise~\cite{Goetz_2016b}. To apply a static magnetic flux to the SQUID loops of the qubits, we use a custom-made superconducting coil with approximately 200 windings.\medskip

\begin{figure}[b]
 \centering
 \includegraphics{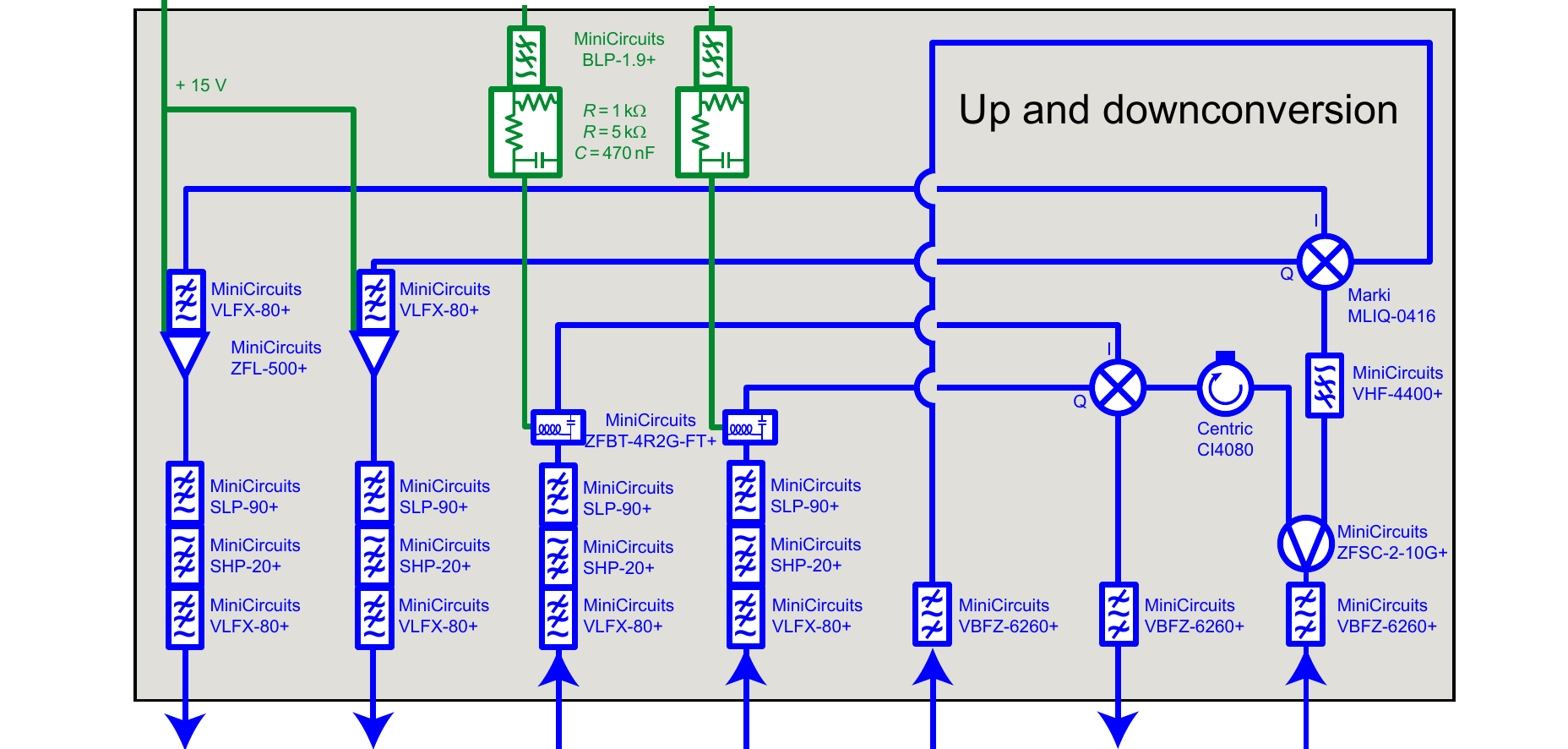}
 \caption{Microwave setup for up and down conversion as well as filtering: We split the coherent signal of a microwave source and feed it to two $\mathcal{IQ}$-mixers for up- and down-conversion. The output of the down-converted signal is amplified before feeding it to the ADCs. We optimize the mixer isolation using a dc voltage fed through bias-tees (green components).}
 \label{fig:Fig_S02a}
\end{figure}

\textbf{Readout setup}---The readout setup that we use for qubit measurements is depicted as the blue components in Fig.\,\ref{fig:Fig_S02} and Fig.\,\ref{fig:Fig_S02a}. Our measurements are carried out with a heterodyne microwave detection setup that is synchronized with the qubit control hardware. We generate coherent readout pulses by up-conversion of a local oscillator signal using a double-balanced $\mathcal{IQ}$ mixer. The temporal envelopes of the readout pulses are rectangular and additionally bandpass filtered. They are generated by a field-programmable gate array (FPGA). The FPGA logic controls the output of two digital-to-analog converters (DACs) running at \SI{250}{\mega\siemens\per\second}. They generate a \SI{62.5}{\mega\hertz} intermediate frequency superimposed to the temporal envelopes of the in-phase and quadrature components, $\mathcal{I}$ and $\mathcal{Q}$, of the readout signal. The pulses are sent through the cryostat, amplified, and down-converted using another $\mathcal{IQ}$ mixer. Finally, we digitize the transmitted $\mathcal{I}$ and $\mathcal{Q}$ components using two analog-to-digital converters (ADCs). These ADCs run at \SI{500}{\mega\siemens\per\second} and are controlled by the same FPGA board as the DACs. Our FPGA code allows us to either analyze single measurement events or ensemble averages to reproduce the time traces in phase space. Currently, the speed of our detection setup is limited by the \SI{40}{\mega\hertz} bandwidth of the band-pass filters in the amplification chain. In addition to the cryogenic amplifiers, this chain contains two room temperature RF amplifiers and a single amplifier for each quadrature component after down-conversion.\medskip

\textbf{Multi-channel readout}---In our experiments, we can apply a readout pulse either to the resonator through the common transmission line and simultaneously to the qubit through a voltage gate line. To ensure the required phase coherence between the two drives, we use a single local oscillator and split the signal after it has passed the $\mathcal{IQ}$ mixer. We obtain full control over the absolute amplitudes of each signal part by using tunable attenuators and digital phase shifters. To optimize the readout performance, we carefully adjust the electrical delay in the different control lines to ensure that the pulses arrive at the sample at identical times.\medskip

\textbf{Qubit control}---All components required for qubit control are depicted in pink in Fig.\,\ref{fig:Fig_S02}. We use a timing module to synchronize the readout FPGA with another FPGA board that generates the qubit control pulses. All these boards are PXI-based and installed in a common controller chassis. The FPGA board for qubit control feeds Gaussian envelopes to the two outputs of a DAC operating at \SI{1.25}{\giga\siemens\per\second}. These envelopes have an intermediate frequency of \SI{312.5}{\mega\hertz} and are upconverted to the qubit frequency using a local oscillator and an $\mathcal{IQ}$ mixer. To optimize the spectral distribution of the qubit control and also of the readout pulses, they can be routed to a spectrum analyzer through a microwave multiplexer. For our experiments we use typical pulse lengths of \SI{100}{\nano\second} to perform a $\pi$ rotation of the qubit state on the Bloch sphere. The FPGA code for the qubit pulses can generate arbitrary waveforms, which allows us to optimize the gate fidelities using randomized benchmarking.

\section{Sample characterization}

\begin{figure}[b]
 \centering
 \includegraphics{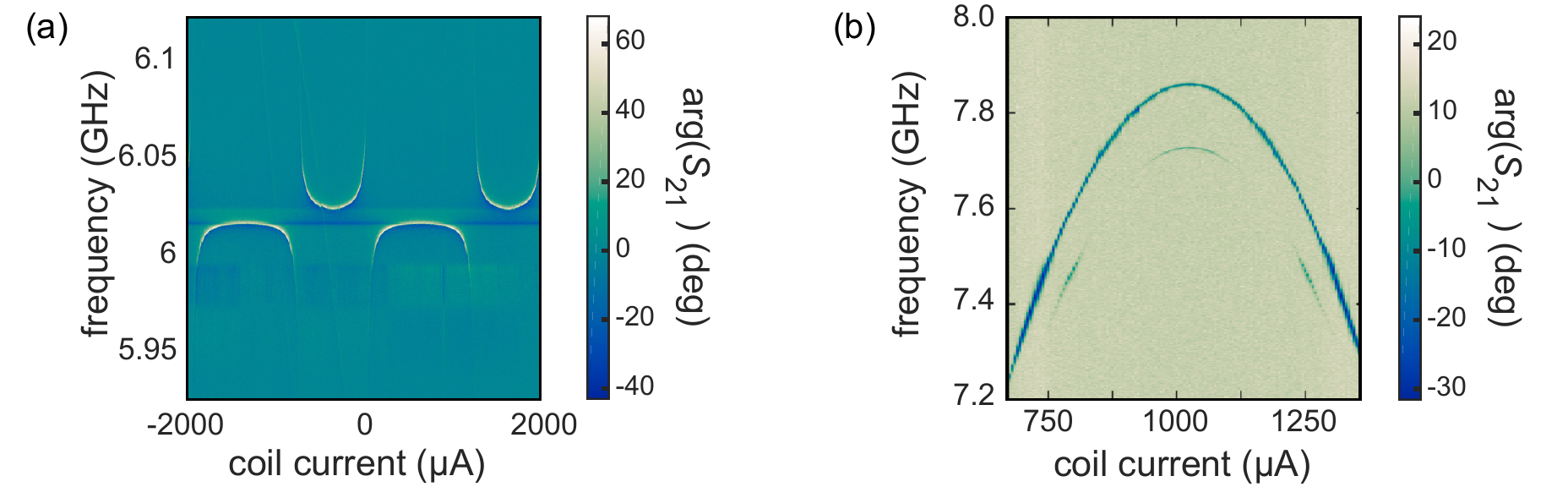}
 \caption{(a) Resonator spectroscopy: Phase response of a readout tone as a function of the probe frequency and the magnetic flux generated by a coil current. We observe clear anticrossings due to the strong coupling between the resonator mode and the transmon qubit. (b) Two-tone qubit spectroscopy: The phase response of the readout tone as a function of the magnetic flux and the frequency of a drive tone applied to the qubit. Due to the strong driving, we observe signatures of the two photon process of the $ |\mathrm{g}\rangle{\,\leftrightarrow\,}|\mathrm{f}\rangle$ transition at frequencies slightly below the qubit $|\mathrm{g}\rangle\,{\leftrightarrow}\,|\mathrm{e}\rangle$ transition.}
 \label{fig:Fig_S03}
\end{figure}

We use spectroscopy measurements to characterize the relevant sample parameters summarized in Table\,\ref{tab:01}. We obtain the resonator frequency and the resonator loss rates from spectroscopy measurements as shown in Fig.\,\ref{fig:Fig_S03}(a). More precisely, we fit an input-output model~\cite{Megrant_2012} to the complex transmission coefficient $S_{21}$ after adjusting the magnetic flux to a value corresponding to the maximum qubit frequency. We additionally extract the qubit--resonator coupling strength $g$ from the anticrossings shown in Fig.\,\ref{fig:Fig_S03}(a). We apply a two-tone measurement protocol to determine the flux-dependent qubit transition frequency $\omega_{\mathrm{q}}(\Phi)\,{=}\,(\sqrt{8E_{\mathrm{J}}E_{\mathrm{c}}|\cos{\pi\Phi/\Phi_{0}}|}\,{-}\,E_{\mathrm{c}})/h$, where $\Phi_{0}$ denotes the flux quantum.
As shown in Fig.\,\ref{fig:Fig_S03}(b), the resonance experiences a frequency shift~\cite{Koch_2007} $\chi/2\pi\,{\equiv}\,g^{2}\alpha/[\Delta(\Delta\,{+}\,\alpha)]\,{=}\,\SI{-1.5}{\mega\hertz}$ depending on the qubit state. This predicted value fits nicely to our experimental result $\chi/2\pi\,{=}\,\SI{-1.6}{\mega\hertz}$, which we extract from measuring the resonator frequency after preparing the qubit either in $\ket{\mathrm{g}}$ or in $\ket{\mathrm{e}}$. We use a high-power drive in these characterization measurements to determine the qubit anharmonicity $\alpha\,{\equiv}\,{-}E_{\text{c}}/h$. The enhanced drive power also excites the two photon process of the $\ket{\mathrm{g}}\,{\leftrightarrow}\,\ket{\mathrm{f}}$ transition, which is detuned from the $\ket{\mathrm{g}}\,{\leftrightarrow}\,\ket{\mathrm{e}}$ transition by $E_{\mathrm{c}}/2h$.

\begin{table}[t]
\caption{\label{tab:01} Overview of the most important sample parameters.}
\begin{tabular}{lr}
\textbf{Qubit parameters} & \\
\hline
transition frequency & $\omega_{\mathrm{q}}/2\pi\,{=}\,\SI{7.86}{\giga\hertz}$\\
charging energy (anharmonicity) & $E_{\mathrm{c}}\,{=}\,h\,{\times}\,\SI{264}{\mega\hertz}$\\
Josephson energy & $E_{\mathrm{J}}\,{=}\,h\,{\times}\,\SI{34}{\giga\hertz}$\\
energy decay rate & $\gamma_{1}\,{=}\,1/(\SI{3.5}{\micro\second})$\\
Ramsey decay rate & $\gamma_{2\mathrm{,R}}\,{=}\,1/(\SI{3.0}{\micro\second})$\\
effective qubit temperature & $T_{\mathrm{eff}}\,{=}\,\SI{73}{\milli\kelvin}$\\
 &\\
qubit--resonator coupling strength & $g/2\pi\,{=}\,\SI{130}{\mega\hertz}$\\
 &\\
\textbf{Resonator parameters}  & \\
\hline
resonance frequency & $\omega_{\mathrm{r}}\,{=}\,2\pi\times\SI{6.02}{\giga\hertz}$\\
external loss rate & $\kappa_{\mathrm{x}}\,{=}\,2\pi\times\SI{1.5}{\mega\hertz}$\\
internal loss rate & $\kappa_{\mathrm{i}}\,{=}\,2\pi\times\SI{0.5}{\mega\hertz}$\\
dispersive shift & $\chi\,{=}\,-2\pi\times\SI{1.6}{\mega\hertz}$\\
\end{tabular}
\end{table}

\newpage
\section{Gate errors and thermal excitations}

\begin{figure}[h]
 \centering
 \includegraphics{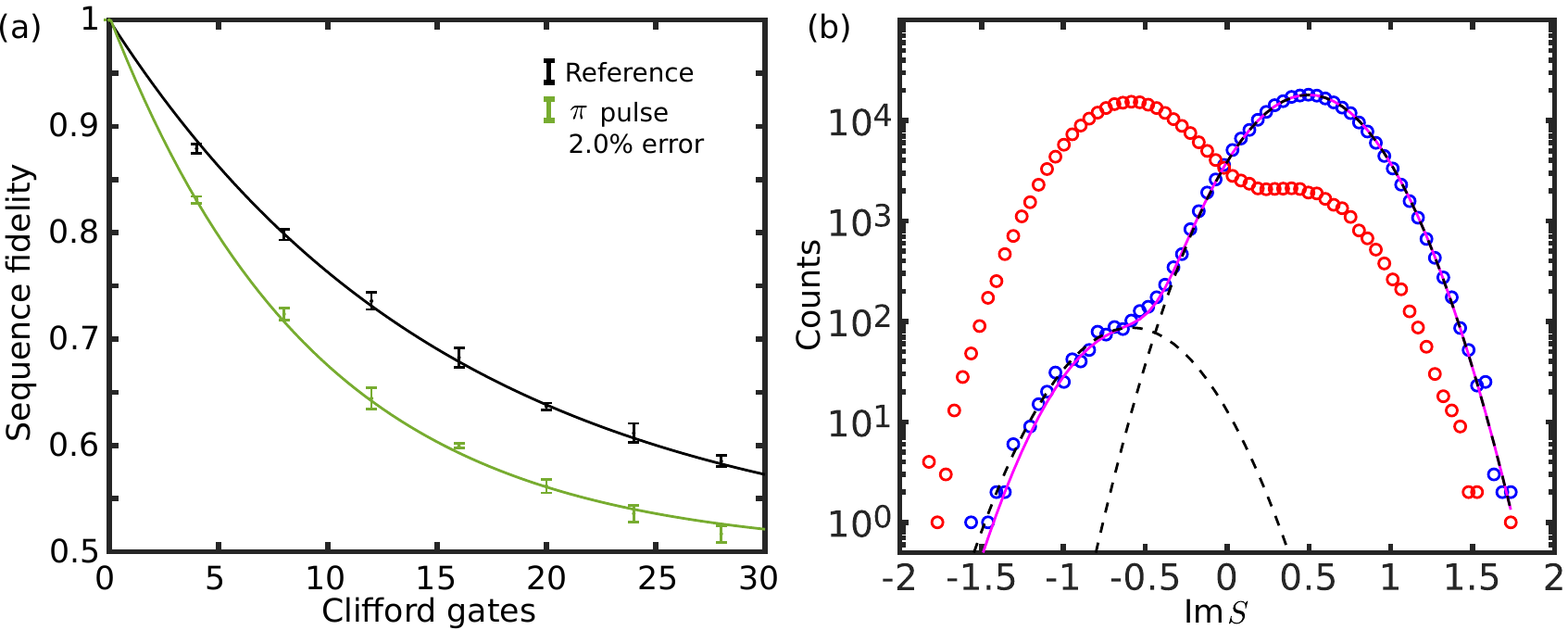}
 \caption{(a) Randomized benchmarking of the gate error: We measure the probability to find the qubit in the ground state after applying a randomized sequence of Clifford gates as a function of the sequence length. For each data point, 80 randomly constructed sequences are applied and measured 4000 times. Repeating the process 3 times yields the averages and standard deviations shown by the markers. (b) Thermal excitations of the qubit: Histogram of $\num{5e5}$ single-shot measurements, integrated without weighting and projected to the imaginary axis. The red and blue circles correspond to qubit prepared in the excited or ground state, repsectively. The latter histogram is a sum of two Gaussian distributions (dashed lines), where the smaller Gaussian represents spontaneous thermal excitations from the qubit ground state. The probability of thermal excitation is obtained by fitting Eq.~\eqref{eqn:gauss} to the ground-state data.}
 \label{fig:Fig_S04}
\end{figure}

\textbf{Gate error}---We implement the randomized benchmarking protocol~\cite{Magesan_2012} to estimate the error caused by non-ideal gate operations when preparing the qubit in the excited state. The protocol amounts to applying a series of quantum gates to the qubit, followed by a standard dispersive readout. The sequence of $L$ gates is built from randomly selected Clifford gates and is terminated with a single gate such that the total operator corresponding to the whole sequence is the identity operator. To estimate the error of a particular gate, in our case the $\pi$ pulse, the selected gate is inserted to the sequence after each random gate. The benchmarking results are shown in Fig.~\ref{fig:Fig_S04}. We fit a function of the form $Ap^L+B$ to the decaying curves of non-interleaved and $\pi$-interleaved sequences to extract the parameters $p_\textrm{ref}$ and $p_\pi$, respectively. The gate error of the $\pi$ pulse is found to be $\epsilon_\textrm{gate}{\,=\,}(1-p_\pi/p_\textrm{ref})/2{\,=\,}(2.0\pm0.3)\%$. Following Ref.~\cite{Epstein_2014}, we further estimate that spontaneous decay during the $\SI{100}{\nano\second}$ gate time contributes with \SI{1.1}{\percent} to $\epsilon_\text{gate}$.\medskip

\textbf{Thermal excitations}---Another source of state preparation error is the thermal excitation of the qubit, the probability of which we extract from the single-shot histogram shown in Fig.~\ref{fig:Fig_S04}(b). We model the single-shot distribution corresponding to the qubit ground state as sum of two Gaussians,
\begin{equation}
Q_\text{g}(z) = (1-\epsilon_\text{th})\exp{-\frac{(z-\alpha_\text{g})^2}{2\sigma_\textrm{th}^2}} + \epsilon_\text{th}\exp{-\frac{(z-\alpha_\text{e})^2}{2\sigma_\textrm{th}^2}}\,,
\label{eqn:gauss}
\end{equation}
where $\epsilon_\text{th}$ denotes the excitation probability and $\sigma_\textrm{th}^2$ is the variance. By fitting this model to the data, we obtain $\epsilon_\text{th}{\,=\,}\SI{0.6}{\percent}$, which is used to calculate the effective qubit temperature as $T_\text{eff}\,{=}\,\hbar\omega_\text{q}/[k_\text{B} \log(1/\epsilon_\text{th})]\,{=}\,\SI{73}{\milli\kelvin}$. Thus the overall state preparation error is $\epsilon_\textrm{prep}\,{=}\,\epsilon_\textrm{gate}\,{+}\,\epsilon_\textrm{th}=2.6\%$, as stated in the main text. We also use the fitted variance $\sigma_\textrm{th}^2$ to obtain a conversion factor from Volts to units of photons for the readout signal. This follows from the fact that the vacuum state is a Husimi distribution with a variance of $1/2$ photons after accounting for the noise of the amplifier.

\section{System Hamiltonian}

Here, we mathematically derive the complete dispersive system Hamiltonian, starting from the system Hamiltonian in the lab frame, we first switch to a rotating frame, in which the results above were simulated. Second, we make make the dispersive approximation, which gives a more intuitive understanding of the system and allows for solving the trajectories analytically.\medskip

\textbf{Laboratory frame}---We treat the qubit as an anharmonic oscillator with eigenfrequencies
$\omega_{k}\,{=}\,k\omega_{\mathrm{r}}\,{+}\,\Delta_{k}$, where $\Delta_{k}$
denotes the detuning between the $k^{th}$ energy level of the qubit
from the resonator angular frequency $\omega_{\mathrm{r}}$. Namely, we define $\Delta_{0}\,{=}\,0$ for the ground state, $\Delta_{1}\,{=}\,\Delta$
for the first excited state, and $\Delta_{2}\,{=}\,2\Delta+\alpha$, where $\alpha$
is the anharmonicity, for the second excited state. In the dispersive
regime, the detuning is larger than the qubit--resonator coupling
strength $g$, i.e., $|\Delta|\,{\gg}\,g$. The Hamiltonian that describes
the system can be written as
\begin{equation}
\hat{H}_{\mathrm{total}}=\hat{H}_{0}+\hat{H}_{\mathrm{int}}+\hat{H}_{\mathrm{QD}}+\hat{H}_{\mathrm{RD}},
\end{equation}
where the free, interaction, qubit-driving, and resonator-driving
Hamiltonians are, respectively, given by 
\begin{eqnarray}
\hat{H}_{0}/\hbar & = & \omega_{\mathrm{r}}\hat{a}^{\dagger}\hat{a}+\sum_{k=0}\omega_{k}\left|k\right\rangle \left\langle k\right|,\\
\hat{H}_{\mathrm{int}}/\hbar & = & \sum_{k=0}g_{k}\left(\hat{a}^{\dagger}+\hat{a}\right)\left(\left|k\right\rangle \left\langle k+1\right|+\left|k+1\right\rangle \left\langle k\right|\right),\\
\hat{H}_{\mathrm{QD}}/\hbar & = & 2\tilde{\Omega}_{\mathrm{q}}(t)\sum_{k=0}\lambda_{k}\left(\left|k\right\rangle \left\langle k+1\right|+\left|k+1\right\rangle \left\langle k\right|\right),\\
\hat{H}_{\mathrm{RD}}/\hbar & = & 2i\tilde{\Omega}_{\mathrm{r}}(t)\left(\hat{a}^{\dagger}-\hat{a}\right).
\end{eqnarray}
Here, $\hat{a}$ denotes the annihilation operator of the resonator
mode, and $\left|k\right\rangle$ refers to
the $k^{th}$ eigenstate of the qubit. For a transmon qubit, the coupling constants
for different transmon levels are typically assumed to be of the form
$g_{k}\,{=}\,g\sqrt{k+1}$, $\lambda_{k}\,{=}\,\sqrt{k\,{+}\,1}$. The real-valued driving waveforms $\tilde{\Omega}_{\mathrm{r/q}}(t)$ at drive frequency $\omega_{\mathrm{d}}$
are constructed from the real and imaginary parts (i.e.,~$\mathcal{I}$ and $\mathcal{Q}$
quadratures) of the complex amplitudes as
\begin{equation}
\tilde{\Omega}_{\mathrm{r/q}}(t)=\mathrm{Re}\left(\Omega_{\mathrm{r/q}}\right)\cos\left(\omega_{\mathrm{d}}t\right)+\mathrm{Im}\left(\Omega_{\mathrm{r/q}}\right)\sin\left(\omega_{\mathrm{d}}t\right).
\end{equation}
\medskip{}

\begin{figure}[h!]
 \centering
 \includegraphics{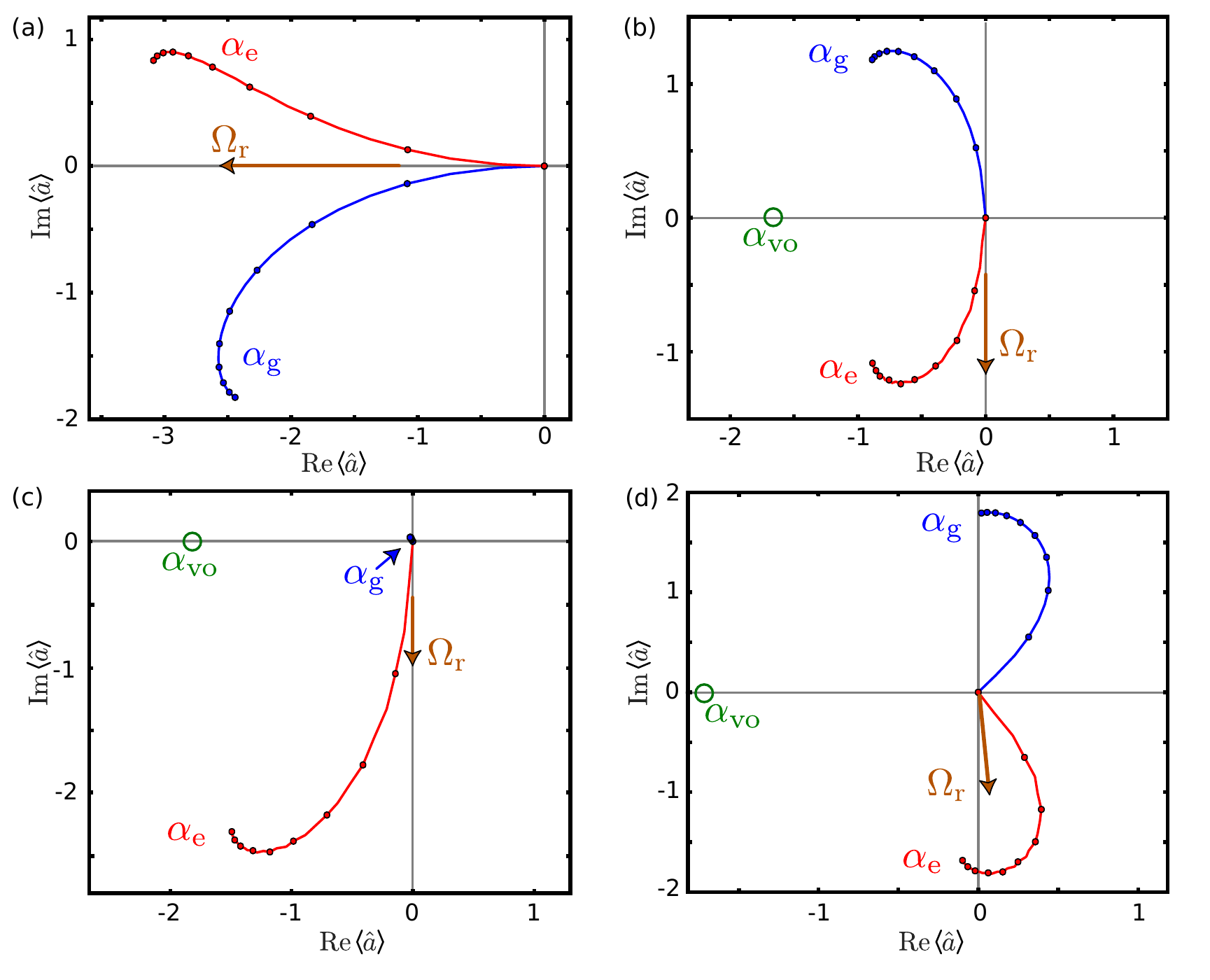}
 \caption{Simulated expectation values $\alpha_\textrm{g/e}=\langle\hat{a}\rangle_\textrm{g/e}$ of the resonator states during a 280-ns-long measurement of the state of a transmon qubit in different readout schemes. We simulate the resonator mode and the qubit as 30 linear and 4 nonlinear energy levels, respectively, with numerical parameters corresponding to those of the experimental sample given in Table~\ref{tab:01}. The drive $\Omega_\text{r}$ applied to the resonator is indicated by the brown arrow and the virtual origin $\alpha_\textrm{vo}=-\Omega_\textrm{q}/g $ due to qubit driving by the green circle. The markers are spaced every \SI{30}{\nano\second}. (a) Typical dispersive readout where driving is only applied to the resonator. (b) Multichannel readout where driving is simultaneously applied to both the qubit and the resonator.  The relative phase of the drives is chosen as indicated in the figure to maximize the initial rate of separation. (c) Multichannel readout where the amplitude of $\Omega_\textrm{r}$ is increased to lock $\alpha_\textrm{g}$ to the origin. (d) As (b), but the  phase of $\Omega_\text{r}$ is changed by $0.1$~rad to show the effect of possibly imprecise phase matching.}
 \label{fig:Fig_S05}
\end{figure}

\textbf{Rotating frame}---We transform $\hat{H}_{\mathrm{total}}$ into the frame rotating at the angular frequency $\omega_{\mathrm{d}}$. Applying the unitary operator 
\begin{equation}
\hat{U}_{1}=\exp\left[it\left(\omega_{\mathrm{d}}\hat{a}^{\dagger}\hat{a}+\sum_{k}k\omega_{\mathrm{d}}\left|k\right\rangle \left\langle k\right|\right)\right]=\mathrm{e}^{it\omega_{\mathrm{d}}\hat{a}^{\dagger}\hat{a}}\sum_{k}e^{it\omega_{\mathrm{d}}k}\left|k\right\rangle \left\langle k\right|,
\end{equation}
and employing the rotating-wave approximations, justified by $g\,{\ll}\,\left|2\omega_{\mathrm{r}}\right|$
and $\left|\omega_{\mathrm{r}}\,{-}\,\omega_{\mathrm{d}}\right|\,{\ll}\,\left|\omega_{\mathrm{r}}\,{+}\,\omega_{\mathrm{d}}\right|$,
yields
\begin{eqnarray}
\hat{H}_{0}^{\prime}/\hbar & = & \hat{U}_1\hat{H}_0\hat{U}_1^\dagger/\hbar + i \dot{\hat{U}}_1\hat{U}_1^\dagger = \left(\omega_{\mathrm{r}}-\omega_{\mathrm{d}}\right)\hat{a}^{\dagger}\hat{a}+\sum_{k}\tilde{\Delta}_{k}\left|k\right\rangle \left\langle k\right|,\label{eq:rotating1}\\
\hat{H}_{\mathrm{int}}^{\prime}/\hbar & = & \hat{U}_1\hat{H}_\text{int}\hat{U}_1^\dagger/\hbar \approx \sum_{k}g_{k}\hat{a}^{\dagger}\left|k\right\rangle \left\langle k+1\right|+\mathrm{H.c.},\\
\hat{H}_{\mathrm{QD}}^{\prime}/\hbar & = & \hat{U}_1\hat{H}_\text{QD}\hat{U}_1^\dagger/\hbar \approx  \Omega_{\mathrm{q}}\sum_{k}\lambda_{k}\left|k+1\right\rangle \left\langle k\right|+\mathrm{H.c.},\\
\hat{H}_{\mathrm{RD}}^{\prime}/\hbar & = & \hat{U}_1\hat{H}_\text{RD} \hat{U}_1^\dagger/\hbar \approx i\Omega_{\mathrm{r}}\hat{a}^{\dagger}+\mathrm{H.c.},\label{eq:rotating4}
\end{eqnarray}
where $\tilde{\Delta}_k\,{=}\,\Delta_{k}\,{+}\,k(\omega_{\mathrm{r}}\,{-}\,\omega_{\mathrm{d}})=\omega_k-k\omega_\text{d}$ denotes the shifted detunings. The total transformed Hamiltonian is given by
\begin{equation}
\hat{H}_{\mathrm{total}}^{\prime}\approx\left(\omega_{\mathrm{r}}-\omega_{\mathrm{d}}\right)\hat{a}^{\dagger}\hat{a}+\sum_{k}\tilde{\Delta}_{k}\left|k\right\rangle \left\langle k\right|+\left\{i\Omega_{\mathrm{r}}\hat{a}^{\dagger} + \sum_{k}\left[g_{k}\hat{a}^{\dagger}+\Omega_{\mathrm{q}}\lambda_{k}\right]\left|k+1\right\rangle \left\langle k\right|+\mathrm{H.c.}\right\}.
\end{equation}

To account for the decay of the resonator state, we use the Lindblad master equation 
\begin{equation}
\ensuremath{\dot{\rho}=-i[\hat{H}_{\text{total}}^{\prime},\rho]/\hbar+\frac{\kappa}{2}\mathcal{L}[\hat{a}]\rho},\label{eq:lindblad}
\end{equation}
where $\rho$ is the reduced density operator of the resonator, $\kappa$ denotes the resonator energy decay rate, and $\mathcal{L}[\hat{a}]\rho=\hat{a}\rho\hat{a}^{\dagger}-\frac{1}{2}\left(\hat{a}^{\dagger}\hat{a}\rho+\rho\hat{a}^{\dagger}\hat{a}\right)$.

We simulate single- and multi-channel readout processes based on driving the resonator and/or the qubit by numerically solving the the master equation with the experimentally obtained parameter values listed in Table~\ref{tab:01}. The results are shown in Fig.~\ref{fig:Fig_S05}. Notably, the figure showcases the different rates at which the states separate in the two readout schemes before saturating towards steady states due to the dissipation. Figure~\ref{fig:Fig_S05} demonstrates that given the experimental parameters and proper calibration of the delays, frequencies, powers, and phases of the drive tones, it is possible to obtain trajectories closer to the ideal case of Figs.~1(b) and~1(c) of the main article than we demonstrate in Figs.~2(b) and~2(c) of the main article. Such fine calibrations are left for future work since the aim of this work was to propose the readout scheme and to experimentally demonstrate its main working principles.

\medskip{}

\textbf{Dispersive approximation}---To make this observation more evident,
we introduce the standard dispersive approximation. We begin by applying
another transformation using
\begin{equation}
\hat{U}_{2}=\exp\left[\sum_{k}\frac{g_{k}}{\tilde{\Delta}_{k+1}-\tilde{\Delta}_{k}}\left(\hat{a}\left|k+1\right\rangle \left\langle k\right|-\hat{a}^{\dagger}\left|k\right\rangle \left\langle k+1\right|\right)\right].
\end{equation}
We compute $\hat{H}_{i}^{\prime\prime}=\hat{U}_{2}\hat{H}_{i}^{\prime}\hat{U}_{2}^{\dagger}$
up to second order in $g_{k}/\tilde{\Delta}_{k+1}$ under the assumption $g_{k}\ll\tilde{\Delta}_{k+1},\forall k$.
For clarity, we restrict the following equations to only the first
three levels of the transmon ($\left\{ \left|\mathrm{g}\right\rangle ,\left|\mathrm{e}\right\rangle ,\left|\mathrm{f}\right\rangle \right\} \equiv\left\{ \left|0\right\rangle ,\left|1\right\rangle ,\left|2\right\rangle \right\} $).
The Hamiltonians become
\begin{eqnarray}
\left(\hat{H}_{0}^{\prime\prime}+\hat{H}_{\mathrm{int}}^{\prime\prime}\right)/\hbar & \approx & \left(\tilde{\Delta}_{1}+\chi_{0}\right)\left|\mathrm{e}\right\rangle \left\langle \mathrm{e}\right|+\left(\tilde{\Delta}_{2}+\chi_{1}\right)\left|\mathrm{f}\right\rangle \left\langle \mathrm{f}\right|\nonumber \\
 &  & +\left[\omega_{\mathrm{r}}-\omega_{\mathrm{d}}-\chi_{0}\left|\mathrm{g}\right\rangle \left\langle \mathrm{g}\right|+\left(\chi_{0}-\chi_{1}\right)\left|\mathrm{e}\right\rangle \left\langle \mathrm{e}\right|+\chi_{1}\left|\mathrm{f}\right\rangle \left\langle \mathrm{f}\right|\right]\hat{a}^{\dagger}\hat{a},\label{eq:H0Hint}\\
\hat{H}_{\mathrm{QD}}^{\prime\prime}/\hbar & \approx & +\Omega_\text{q} \left|\mathrm{e}\right\rangle \left\langle \mathrm{g}\right| + \Omega_\text{q} \left|\mathrm{f}\right\rangle \left\langle \mathrm{e}\right|\nonumber \\
 &  & -\Omega_\text{q}\frac{\chi_0}{g_0}\hat{a}^\dagger\left|\mathrm{g}\right\rangle \left\langle \mathrm{g}\right|+\frac{\chi_{0}-\chi_{1}}{g_0}\Omega_{\text{q}}\hat{a}^\dagger\left|\mathrm{e}\right\rangle \left\langle \mathrm{e}\right|+\Omega_{\text{q}}\frac{\chi_1}{g_0}\hat{a}^\dagger\left|\mathrm{f}\right\rangle \left\langle \mathrm{f}\right|+\mathrm{H.c.},\label{eq:HQD}\\
\hat{H}_{\mathrm{RD}}^{\prime\prime}/\hbar & \approx & i\Omega_{\mathrm{r}}\left(\hat{a}^{\dagger}+\frac{\chi_{0}}{g_{0}}\left|\mathrm{e}\right\rangle \left\langle \mathrm{g}\right|+\frac{\chi_{1}}{g_{1}}\left|\mathrm{f}\right\rangle \left\langle \mathrm{e}\right|\right)+\mathrm{H.c.},\label{eq:HRD}
\end{eqnarray}
where we have defined the dispersive constants $\chi_{0}=g_{0}^{2}/\tilde{\Delta}_{1}$
and $\chi_{1}=g_{1}^{2}/(\tilde{\Delta}_{2}-\tilde{\Delta}_{1})$.

Finally, introducing the displaced operator $\hat{b}=\hat{a}-\alpha_{\text{vo}}$ where $\alpha_{\text{vo}}\equiv\,{-}\Omega_{\mathrm{q}}/g$,
the total Hamiltonian reads 
\begin{eqnarray}
\hat{H}_\textrm{total}^{\prime\prime}/\hbar & \approx & \chi_{0}\left|\alpha_{\text{vo}}\right|^{2}\left|\mathrm{g}\right\rangle \left\langle \mathrm{g}\right|+\left[\tilde{\Delta}_{1}+\chi_{0}-\left|\alpha_{\text{vo}}\right|^{2}\left(\chi_{0}-\chi_{1}\right)\right]\left|\mathrm{e}\right\rangle \left\langle \mathrm{e}\right|+\left[\tilde{\Delta}_{2}+\chi_{1}\left(1-\left|\alpha_{\text{vo}}\right|^{2}\right)\right]\left|\mathrm{f}\right\rangle \left\langle \mathrm{f}\right|\label{eq:row1}\\
 &  & +\left[\left(-\alpha_\text{vo}g_{0}+\Omega_{\mathrm{r}}\frac{\chi_{0}}{g_{0}}\right)\left|\mathrm{e}\right\rangle \left\langle \mathrm{g}\right|+\left(-\alpha_\text{vo}g_{1}+\Omega_{\mathrm{r}}\frac{\chi_{1}}{g_{1}}\right)\left|\mathrm{f}\right\rangle \left\langle \mathrm{e}\right|+\mathrm{H.c.}\right]\label{eq:row2}\\
 &  & +\left[\omega_{\mathrm{r}}-\omega_{\mathrm{d}}-\chi_{0}\left|\mathrm{g}\right\rangle \left\langle \mathrm{g}\right|+\left(\chi_{0}-\chi_{1}\right)\left|\mathrm{e}\right\rangle \left\langle \mathrm{e}\right|+\chi_{1}\left|\mathrm{f}\right\rangle \left\langle \mathrm{f}\right|\right]\hat{b}^{\dagger}\hat{b}\label{eq:row3-1}\\
 &  & +\left[i\Omega_{\mathrm{r}} + \alpha_\textrm{vo}(\omega_{\mathrm{r}}-\omega_{\mathrm{d}})\right]\hat{b}^{\dagger}+\text{H.c.}\label{eq:row4}
\end{eqnarray}

Line~(\ref{eq:row1}) describes the constant frequency shifts caused
by the coupling and the driving. Line~(\ref{eq:row2}) shows that
driving from the qubit side tilts the qubit Hamiltonian. Importantly, line~(\ref{eq:row3-1})
predicts that any coherent state will rotate about point $\alpha_{\text{vo}}$. The angular frequencies of these rotations may be set to be equal to $+\chi\equiv\chi_{0}\,{-}\,\chi_{1}/2$ and ${-}\chi$ for $\alpha_\textrm{e}$ and $\alpha_\textrm{g}$, respectively, by choosing  $\omega_\textrm{r}\,{-}\,\omega_\textrm{d}\,{=}\,\chi_1/2$. Note that non-linearity is important here.
Line~(\ref{eq:row4}) shows that the transformation has an effect on the amplitude of the resonator drive that may be compensated by changing $\Omega_\textrm{r}$. Note that the Hamiltonian of a conventional dispersive system is obtained by setting $\alpha_{\text{vo}}\,{=}\,0$.

The two-level Hamiltonian given by Eq.~(2) in the manuscript is produced from Eqs.~\eqref{eq:H0Hint}--\eqref{eq:HRD} by assuming that the third level of the transmon is never populated, setting the corresponding couplings to zero, $g_{1}\,{=}\,\chi_{1}\,{=}\,0$, and making the above-discussed choice  for $\omega_\text{d}$. For Eq.~(2), we have re-labeled $g_{0}\rightarrow g$.

\medskip{}

\textbf{Trajectories and steady states}---Using Eq.~\eqref{eq:lindblad} with the approximate Hamiltonian $\hat{H}^{\prime\prime}_\textrm{total}$, we obtain an analytical equation for the expectation value $\alpha_{j}=\bra{\alpha_j}\hat{a}\ket{\alpha_j}$, $j\in\{\text{g,e}\}$, as
\begin{equation}
\frac{\partial\alpha_{\text{g/e}}(t)}{\partial t}=\Omega_{\mathrm{r}}\pm i\chi\left[\alpha_{\text{g/e}}(t)-\alpha_{\text{vo}}\right]-\frac{\kappa}{2}\alpha_{\text{g/e}}(t),
\end{equation}
Choosing constant control pulses, $\partial\alpha_{\text{vo}}/\partial t\,{=}\,0$, and assuming that
the resonator is initially in the vacuum state, the solution is given by
\begin{eqnarray}
\alpha_{\text{g/e}}(t) & = & \frac{i\Omega_{\mathrm{r}}\mp\Omega_{\mathrm{r}}\chi/g}{i\kappa/2\pm\chi}\left[1-\exp\left(\pm it\chi-\frac{\kappa t}{2}\right)\right].
\end{eqnarray}
Setting $t\rightarrow\infty$ results in the steady state formulae given in the manuscript.

\bibliographystyle{apsrev4-1}
\bibliography{Goetz_Bibliography}